\documentclass[aps,prl,superscriptaddress,floatfix,reprint]{revtex4-1}
\usepackage{graphicx}
\usepackage{amsmath}
\usepackage{amssymb}
\usepackage{hyperref}
\usepackage{color}
\usepackage{CJKutf8}

\newcommand{\bmat}{\left ( \begin{array}{cc}}
\newcommand{\emat}{ \end{array}\right )}

\renewcommand\epsilon\varepsilon
\renewcommand\phi\varphi
\renewcommand\Im{{\rm Im}}

\newcommand\be{\begin{eqnarray}}
\newcommand\ee{\end{eqnarray}}
\newcommand{\nn}{\nonumber}
\newcommand\eref[1]{(\ref{#1})}

\newcounter{yjcc}

\begin{document}

\title{\boldmath Replica Symmetry Breaking and Phase Transitions in a PT Symmetric Sachdev-Ye-Kitaev Model}

\author{Antonio M. Garc\'\i a-Garc\'\i a}
\affiliation{Shanghai Center for Complex Physics,
	School of Physics and Astronomy, Shanghai Jiao Tong
	University, Shanghai 200240, China}
\email{amgg@sjtu.edu.cn}
\author{Yiyang Jia\begin{CJK*}{UTF8}{gbsn}
		(贾抑扬)
\end{CJK*}}
\affiliation{Department of Physics and Astronomy, Stony Brook University, Stony Brook, New York 11794, USA}
\email{yiyang.jia@stonybrook.edu}

\author{Dario Rosa}
\affiliation{Center for Theoretical Physics of Complex Systems, Institute for Basic Science(IBS), Daejeon 34126, Korea}
\affiliation{Department of Physics, Korea Advanced Institute of Science and Technology,
291 Daehak-ro, Yuseong-gu, Daejeon 34141, Republic of Korea}
\email{dario\textunderscore rosa@ibs.re.kr}

\author{Jacobus J. M. Verbaarschot}
\affiliation{Department  of Physics and Astronomy, Stony Brook University, Stony Brook, New York 11794, USA}
\email{jacobus.verbaarschot@stonybrook.edu}

\date{\today}

\begin{abstract}
  We show that the low temperature phase of a conjugate pair of uncoupled, quantum chaotic, nonhermitian systems such as the Sachdev-Ye-Kitaev (SYK) model or the Ginibre ensemble of random matrices are dominated by replica symmetry breaking (RSB) configurations with a nearly
  flat free energy that terminates in a first order phase transition. In the case of the SYK model, we show explicitly that the spectrum of the effective replica theory has a gap. These features are strikingly similar to those induced by wormholes in the gravity path integral which suggests a close relation between both configurations. For a non-chaotic SYK, the results are qualitatively different: the spectrum is gapless in the low temperature phase and there is an infinite number of second order phase transitions unrelated to the restoration of replica symmetry.
\end{abstract}

\maketitle

\allowdisplaybreaks[3]


The study of nonhermitian effective Hamiltonians has a long history \cite{bender1998,hatano1996}. Perhaps the best known example is the effective Hamiltonian that describes
resonances with a finite width, for example the one that enters in the
calculation of the $S$-matrix of open quantum systems such as quantum dots
 \cite{alhassid2000} 
or compound nuclei \cite{Verbaarschot:1985jn}.
Another example is the Euclidean QCD Dirac operator at
nonzero chemical potential, which is nonhermitian
with spectral support on a two-dimensional domain of the complex plane \cite{Barbour:1986jf}.
In hermitian theories, a phase transition may arise due to the formation
of a gap. This may also happen for nonhermitian systems when the domain
of eigenvalues splits into two or more pieces. However, another mechanism
is possible. Because of the nonhermiticity, the action
is generally complex, and the saddle point with the largest real part of the
free energy may get nullified after ensemble averaging.
In QCD at nonzero baryon chemical potential, the pion condensation phase
is nullified so that the phase transition to nonzero baryon density becomes
visible \cite{Hands:2007by}. The conclusion is that the phase diagram can be altered dramatically
by the nullification of the leading saddle point \cite{Osborn:2005ss,Hands:2007by}.

A second point we wish to make is about the nature of quenched averages in nonhermitian theories.  Although alternatives are possible \cite{efetov1983,Verbaarschot:1985jn,kanzieper2002}, quenched averages are often carried out by means of the replica trick \cite{edwards1975}. However, because of  Carlson's theorem
\cite{carlson1914}, a naive application of the replica trick is not guaranteed to work \cite{verbaarschot1985}. The best known
example of the failure of the replica trick is in the calculation of the quenched free energy of the Sherrington-Kirpatrick model \cite{sherrington1972}, a toy model for spin glasses, which in the low temperature limit yields a negative entropy \cite{sherrington1972}. This inconsistency was
ultimately resolved by postulating a ground state that breaks the replica
symmetry \cite{parisi1979,mezard1984}. The problems with the replica trick are more
dramatic for nonhermitian theories as was first demonstrated for QCD
at nonzero chemical potential $\mu$ \cite{Stephanov:1996ki}. In this case, the $n$ replica
(or $n$ flavor) partition function is given by
\be
Z_n = \langle {\det}^n D(\mu)\rangle,
\ee
where the averaging is over gauge field configurations weighted by the
Euclidean Yang-Mills action. It was shown that the quenched approximation, where the determinant is put to unity, is not given by
$\lim_{n\to 0}  Z_n$ but rather by
\be
\lim_{n\to 0}   \langle  {\det}^n (D(\mu)D^\dagger(\mu))\rangle.
\ee
Because the disconnected part
of the partition function
is nullified
due to the phase of the fermion determinant, this partition function is
dominated by replica symmetry breaking (RSB) configurations which in this context are referred to as Goldstone bosons of a quark and a conjugate quark. A similar RSB mechanism has been identified in the context of random matrix theory, for both hermitian \cite{mezard1999,okuyama2020} and nonhermitian \cite{nishigaki2002} random matrix ensembles.

The possibility of a partition function where the connected part
dominates the disconnected part has recently received a great deal of
attention in the analysis of wormhole solutions in Jackiw-Teitelboim (JT) \cite{jackiw1985,teitelboim1983} gravity
and related theories \cite{Saad:2018bqo,Saad:2019lba,Saad:2019pqd,Almheiri:2019psf,engelhardt2020,penington2020,almheiri2020,vanraamsdonk2020,garcia2020,salomon2020,turiaci2020}. 
The existence of these solutions in Lorentzian signature was first observed in
\cite{maldacena2018}
with the discovery of a low temperature traversable wormhole phase in a near AdS$_2$ background
deformed by weakly coupling the two boundaries.
\textit{}As temperature increases, the system eventually undergoes a first order wormhole to black hole transition. By adding complex sources, it is possible to find \cite{garcia2020} Euclidean wormholes solutions of JT gravity that undergo a similar transition at finite temperature.

Interestingly, a replica calculation \cite{engelhardt2020} of the quenched free energy in JT gravity found that, in the low temperature limit, the contribution of replica wormholes is dominant. Likewise, the evaluation of the von Neummann entropy by the replica trick \cite{penington2020,almheiri2020,almheiri2020a,vanraamsdonk2020} revealed the existence of additional RSB saddle points, wormholes
connecting different copies of black holes in this context. These wormhole configurations are crucial to make the process of black hole evaporation consistent with unitarity \cite{page1993}.

A natural question to ask is whether
these replica wormholes have a field theory analogue. RSB configurations have indeed been explored in the SYK model with real couplings \cite{arefeva2018,wang2018}. However, there is no evidence that they dominate the partition function.

In this paper, we answer the question posed in the previous paragraph affirmatively
by identifying a pair of nonhermitian random Hamiltonians whose sum is PT-symmetric where RSB configurations are the leading saddle points of the action in the low temperature phase.
We consider a nonhermitian version of the SYK
model and show that it has a phase transition from a phase dominated
by the disconnected part of the partition function to a phase dominated
by the connected part, namely, a phase dominated by RSB configurations.

The $q$-body SYK Hamiltonian \cite{french1970,bohigas1971,sachdev1993,kitaev2015} is
defined by
\be
H _{\rm SYK} = (i)^{q/2} \sum_{\alpha_1< \cdots <\alpha_q} J_{\alpha_1\cdots\alpha_q}
\chi_{\alpha_1}\cdots \chi_{\alpha_q}
\ee
where the $\chi_{\alpha}$ represent $N$ Majorana fermions, satisfying the anti-commutation
relations $\{\chi_\alpha,\chi_\beta\} = \delta_{\alpha \beta} $, and the $J_{\alpha_1\cdots\alpha_q}$ are real couplings, sampled from a Gaussian distribution having a vanishing mean value and a variance proportional to $1/N^3$. The coupled SYK model introduced
by Maldacena and Qi (MQ) 
in \cite{maldacena2018} consists
of a Right (R) SYK model and a Left (L) SYK model each with $N/2$ Majorana fermions, and a coupling term
$i\mu \sum_{k=1}^{N/2} \chi^R_k \chi^L_k$. Although the left and right
couplings, denoted by $J^{L(R)}_{\alpha_1\cdots\alpha_q}$
are chosen to be the same as in the MQ model,  it is also possible, as was noted by the same authors,
to take them different. One remarkable observation was
made: the solution that couples the right and left SYK continues to exist
in absence of an explicit coupling ($\mu =0$) provided that
\be
\langle J^L J^R \rangle > \langle J^L J^L\rangle = \langle J^R J^R \rangle
\ee
where $\langle \ldots \rangle$ stands for ensemble average.
Since the covariance matrix is no longer positive, this cannot be realized
by real-valued $J_L$ and $J_R$. However, this can be
achieved for the complex couplings
\be\label{jljr}
J^L=J+ik K,\qquad 
J^R =J-ik K
\ee
with $J$, $K$ independent real Gaussian stochastic variables with the same variance
and zero mean.

Before continuing, let us analyze the quenched free energy of the single-site SYK we have
just introduced:
\be
\langle \log Z \rangle = \langle \log |Z|\rangle  + i \langle \arg Z\rangle,
\ee
where $Z$ is the partition function for a specific realization of the couplings.
Since the phase of the partition function does not have a preferred
direction, we expect that $ \langle \arg Z\rangle =0$. We conclude
\be
\langle \log Z \rangle = \frac 12 \langle \log Z Z^*\rangle
\ee
for a theory where $Z$ and $Z^*$ have equal probability.
In particular, the quenched free energy is given by the replica limit
\be
\frac 12 \lim_{n\to  0} \left \langle\frac {(Z Z^*)^n -1}n \right \rangle.
  \label{rep-lim}
  \ee
  Therefore, we arrive naturally at a system of two conjugate SYK Hamiltonians
 only coupled through the
  probability distribution. The Hamiltonian corresponding to $Z Z^*$ is given by
  \be
  H_{\rm SYK} \otimes 1 + 1 \otimes H_{SYK}^\dagger ,
\label{ham-pt}
  \ee
  which is exactly the two-site Hamiltonian proposed in \cite{maldacena2018} with the explicit coupling turned off.
This Hamiltonian is $PT$ symmetric \cite{bender1998},
where the $P$ operator interchanges the L and R spaces and
$T$ is the tensor product of two copies of the time reversal operator for the standard SYK model \cite{you2016,garcia2016,Cotler:2016fpe}.

\begin{figure}[t!]
    \includegraphics[width=8cm]{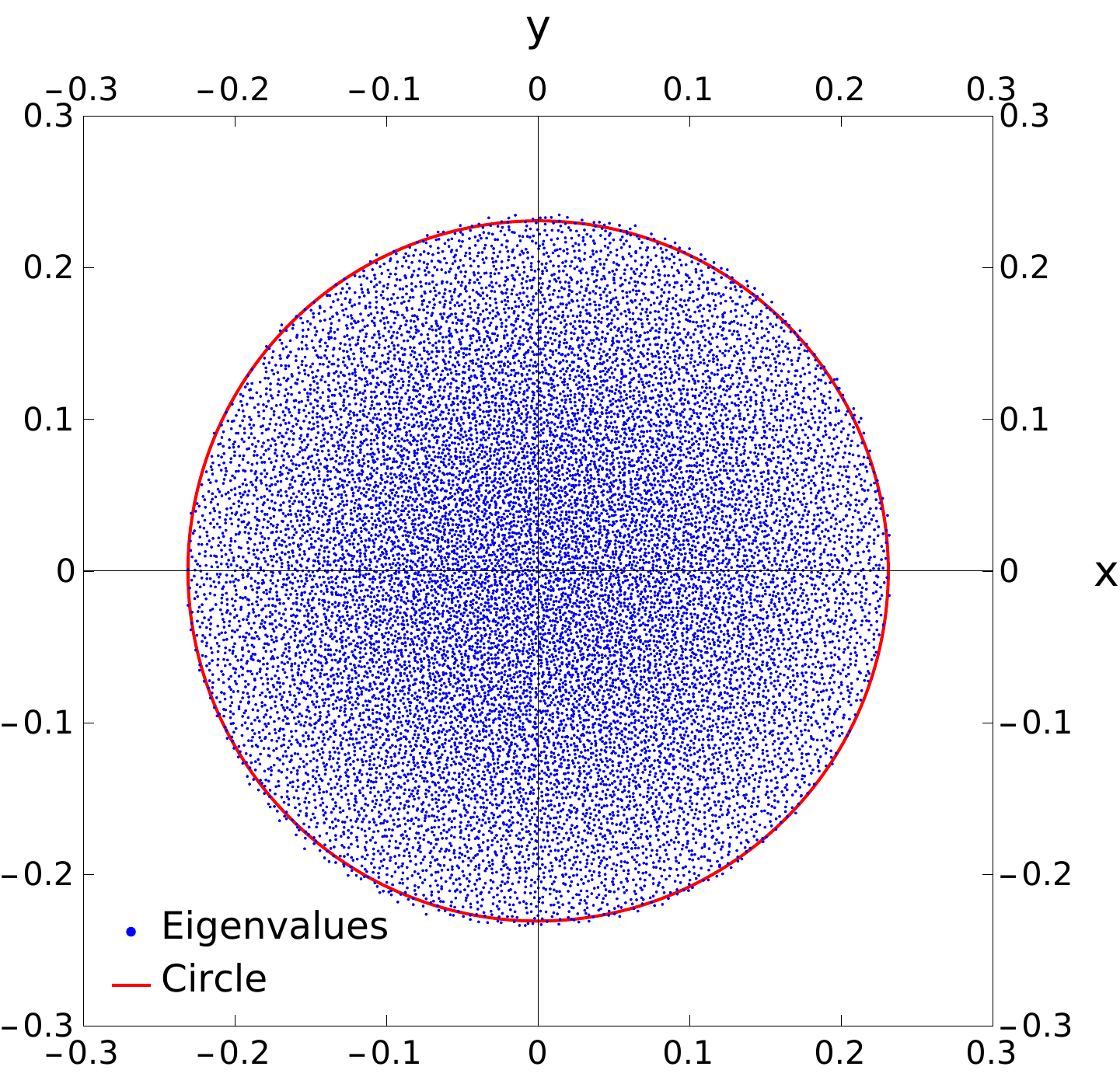}
    \caption{The eigenvalue density, obtained from exact diagonalization, for one realization
      of the $q = 4, k = 1$ nonhermitian SYK model with $N/2=30$,
      compared to a circle (red curve). }
      \label{fig:dens}
    \end{figure}

  Next, we calculate the partition function $\langle Z Z^* \rangle$ for large $N$. We
  expect that in this limit
  \be
  \langle (Z Z^*)^n \rangle  =  
  \langle Z Z^* \rangle ^n 
  \ee
  so that the replica limit \eref{rep-lim} requires only the computation of $ \langle Z Z^* \rangle$.
  We consider a pair of nonhermitian SYK models with couplings~(\ref{jljr}) for $k=1$. The spectral density
  is given by a disk with radius $E_0$ in the complex plane. Although the 
  eigenvalue density is rotationally invariant, it 
  is not constant as is the case for the large $N$ limit of the Ginibre \cite{ginibre1965,fyodorov1997} ensemble or random matrices
  (see Fig. \ref{fig:dens}).
  However, we expect that eigenvalue correlations are in the universality class of
  the Ginibre model.
 If the averaged eigenvalue
  density is denoted
  by $\rho(z)$, the two-level correlation function is given by
  \be
  \rho_2(z_1,z_2) = R_{2c}(z_1,z_2) + \delta^2(z_1-z_2)\rho(z_1)+
  \rho(z_1)\rho(z_2). \nn
  \ee
  where $R_{2c}(z_1,z_2)$ is the averaged connected two-point correlation function not including the  self-correlations.   The partition function is given by
  \be\label{eqn:two-siteParitionFunc}
  \hspace*{-0.5cm}\langle Z Z^* \rangle =
\langle Z \rangle \langle Z^* \rangle +\langle Z Z^* \rangle_c
\ee 
where $ \langle Z Z^* \rangle_c = \int d^2 z_1 d^2 z_2 \rho_{2c}(z_1,z_2)
  e^{-\beta(z_1 +z_2^*)}$, $\rho_{2c}(z_1,z_2) = R_{2c}(z_1,z_2) + \delta^2(z_1-z_2)\rho(z_1)$ and
    $
    \langle Z \rangle = \int d^2z \rho(z) e^{-\beta z}$.
    Because the eigenvalue density has rotational invariance, we can use
    the mean value theorem to show that the partition function is independent
    of $\beta$ and given by the normalization of $\rho(z)$, which we denote by $D$, $
    \langle Z \rangle = D$.

    To evaluate the  second term of \eqref{eqn:two-siteParitionFunc}, we use the sum rule
    \be
    \int d^2 z_2 (R_{2c}(z_1,z_2)
    + \delta^2(z_1-z_2)\rho(z_1))=0,
    \label{sum-rule}
    \ee
    and the fact that the correlations are short-range \cite{ginibre1965,fyodorov1997} with the
    connected correlator taking the universal form
    \be
    R_{2c}(z_1,z_2) = R_{2c}^{\rm unv} (\sqrt{\rho(\bar z)}(z_1-z_2)) \rho(\bar z)^2,
    \ee
    where $\bar z=(z_1+z_2)/2$.
 We  thus have that $|z_1 -z_2|< 1/ \sqrt D$ region gives the dominant contribution
 and we can Taylor expand the exponent in (\ref{eqn:two-siteParitionFunc}) in powers of
 $\beta \Im(z_1 -z_2)$.
    The zero order term vanishes because of the sum rule \eref{sum-rule}, the linear term vanishes because the probability distribution is even under complex conjugation.   After performing the integral over  $z_1- z_2$, we obtain the
connected  partition function
\be
\langle Z Z^* \rangle_c &=&  \beta^2 \langle\zeta^2 \rangle \int_{|\bar z| < E_0} d^2 \bar z
e^{-\beta(\bar z+\bar z^* )}\nn\\
&=&  \pi \beta E_0 \langle\zeta^2 \rangle I_1(2\beta E_0),
\ee
where $\langle \zeta \rangle =\langle (\Im(z_1- z_2))^2\sqrt{\rho(\bar z)}\rangle  \sim D^0$, the ground state energy $E_0 \sim N$ and $D = 2^{N/4}$.
Including the disconnected part of the partition function and using the asymptotic expression of $I_1$, we obtain a free energy
\be
F = -T \log \left [e^{2 \beta E_0} + 2^{N/2}\right ]
\ee
where we have neglected prefactors that are subleading in $N$.
In the strict large $N$ limit, it simplifies to,
\be
\frac{F(T)}N =\frac {-2E_0}N\theta (T_c-T)   - \frac{T\log 2}{2} \theta(T-T_c) ,
\label{ft-an}
\ee
with
$
T_c = {4E_0}/(N \log 2).
$
\begin{figure}[t!]
\centerline{  \includegraphics[width=8cm]{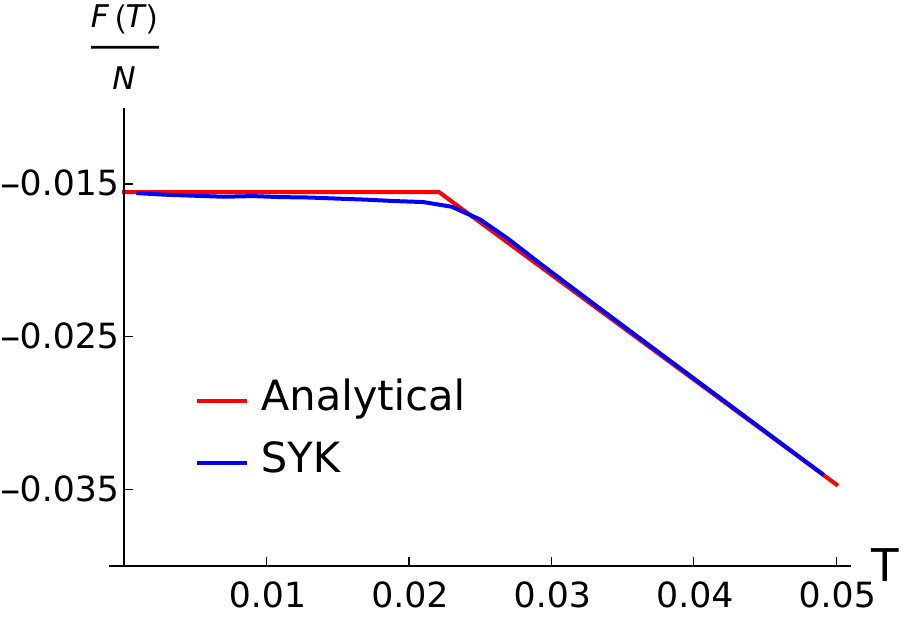}}
  \caption{The temperature dependence of the free energy of the nonhermitian
    SYK model for $N/2 = 30$, $ q=4$ and $k=1$ (blue curve) compared to
    the analytical result \eref{ft-an}.
  The value of $E_0$ is the radius of the circle in Fig. \ref{fig:dens}.}
    \label{fig:free}
\end{figure}
In Fig. \ref{fig:free}, we show the numerical quenched free energy of the SYK
Hamiltonian \eref{ham-pt} for $N/2=30$,  $q=4$ and $k=1$ (black curve)
and compare it to the analytical result \eref{ft-an}. The deviation in the constant part seems to scale as $1/N$. The free energy can also be worked out for
$k<1$, where we also find a first order phase transition with $T_c \sim k^4$
for small $k$.

We have thus observed that the leading exponent of the disconnected part of the partition function is nullified by the phase of the Boltzmann factor so that the contribution due to the connected
part of the two-point correlation function becomes dominant. The free energy
behaves as if the system has a gap. Both features are typical of RSB configurations. 

In order to make this connection more explicit, we show that
these results can also be obtained by solving the Schwinger-Dyson equations, in the $\Sigma G$ formulation of the two-site SYK model 
\cite{maldacena2018,garcia2019} which is equivalent to performing the replica trick and then solving the model in the saddle point approximation.
  For $T> T_c$ and $k = 1$, the solution
with the free $G$ and $\Sigma$ is dominant so that only the kinetic term
of the Lagrangian remains. As a consequence, the free energy is  $-T \log 2/2$ in agreement with the spectral calculation above. For $T<T_c$, a nontrivial RSB
solution becomes dominant which results in a constant free energy 
up to exponentially small corrections. Similar results can be derived for $k<1$, where, in agreement with the previous analytical calculation, we have also found $T_c \sim k^4$. Indeed, this feature is shared by both Euclidean \cite{garcia2020} and traversable \cite{maldacena2018, maldacena2020syk} wormholes. 
Details of this and the previous analytical calculation will be given elsewhere \cite{us2021}.
  
\begin{figure}[t!]
	\centering
  \includegraphics[width=7cm]{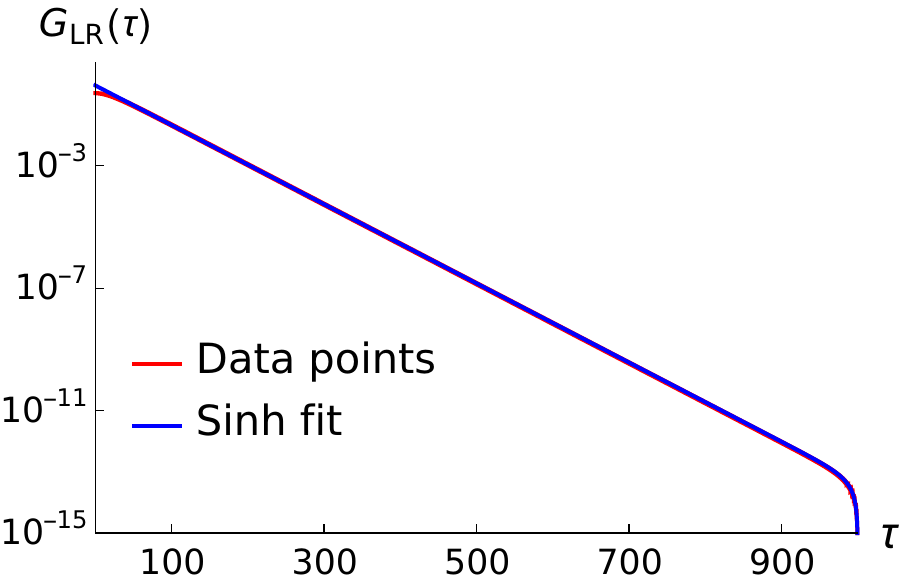} 
	\includegraphics[width=7cm]{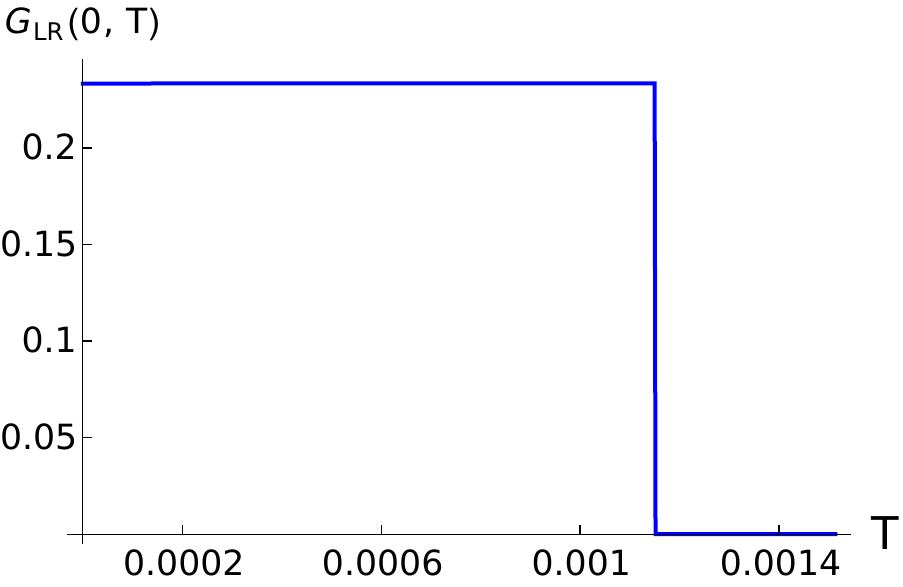}  
	\includegraphics[width=7cm]{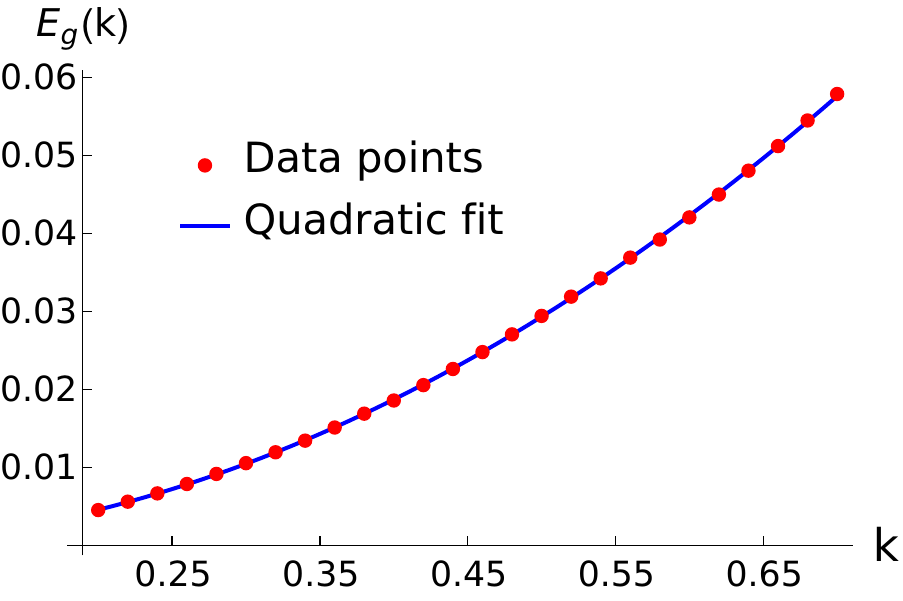} 
	\caption{Top: $G_{LR}(\tau)$ from the solution of the Schwinger-Dyson equations for the SYK model ~(\ref{ham-pt}) with $q=4$, $T = 0.0005$ and $k = 0.5$ fitted by (\ref{eq:mass_gap_ansatz}). Middle: The order parameter
          $G_{LR}(0)$, versus temperature for $k = 0.5$. Bottom: The energy gap $E_g$ for
          $T = 0.0005\ll T_c$ as a function of $k$ from the fit (\ref{eq:mass_gap_ansatz}).
                }
	\label{Fig:energy_gap}
\end{figure}
  For traversable wormholes \cite{maldacena2018}, the spectrum is gapped. Physically, it is related to the interaction-driven tunneling between the left and right sites. 
  The existence of the energy gap can be demonstrated  \cite{maldacena2018}
  directly from the effective boundary gravity action or
  from the exponential decay of the left-right Green's function, $G_{LR}(\tau)$,
of the two-site SYK model 
for low temperatures.  We study whether a similar gap exists in the two-site non-hermitian SYK. We stress the gap in this case is not a property of the microscopic Hamiltonian ~(\ref{ham-pt}) but only of the resulting replica field theory after ensemble average. Since $G_{LR}(\beta/2-\tau) =-G_{LR}(\tau-\beta/2)$  we employ the  ansatz \cite{maldacena2018}
\begin{equation}
\label{eq:mass_gap_ansatz}
G_{LR} (\tau) \sim \sinh E_{g} (\beta/2-\tau) \ ,
\end{equation}
where the gap $E_{g}$ is a fitting parameter. The fit is excellent except for
very small times \cite{Plugge_2020}, see Fig.  ~\ref{Fig:energy_gap}.
This reinforces the picture that RSB configurations mediate tunneling between the two sites even though there is no direct coupling term
in the Hamiltonian. We note that $G_{LL}$ and $G_{RR}$ show a similar decay.
In Fig. \ref{Fig:energy_gap}, we also show that the
gap $E_g$ depends quadratically on $k$.
It would be interesting to understand this exponent from the gravity side. 
We propose $G_{LR}(0)$ as the order parameter of the transition since a non-vanishing $G_{LR}$ is a distinctive feature of RSB configurations. Results depicted in
in Fig. \ref{Fig:energy_gap}, confirm that $G_{LR}(0)$ remains almost constant in the wormhole phase and vanishes for  $T > T_c$.\\
 \begin{figure}[t!]
	\centerline{\includegraphics[width=8cm]{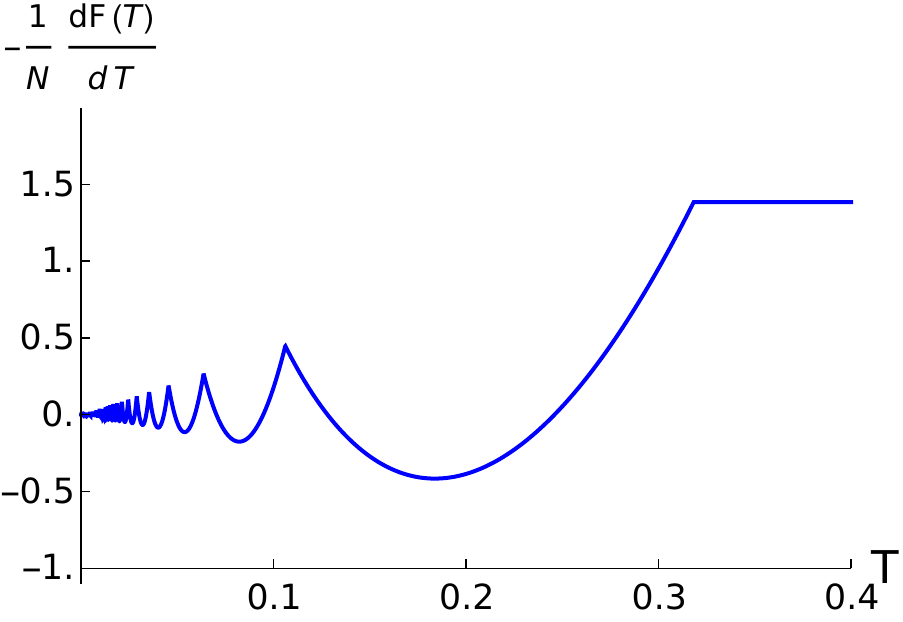}}
	\caption{Derivative of the free energy for the nonhermitian $q=2, k=1$ SYK model. On the way to $T=0$, the system undergoes infinitely
		many second order phase transitions. For $T > 1/\pi$, it becomes a constant which is a typical feature of free fermions.} 
	\label{fig:dfree}
\end{figure}

The studied $q = 4$ SYK model is quantum chaotic \cite{kitaev2015}. We expect very similar results in others quantum chaotic systems such as $q>2$ SYK models and the Ginibre ensemble of random matrices \cite{ginibre1965} because RSB depends on the connected two-level correlation function which, to some extent, is universal in this case \cite{bohigas1984}. 
However, it is unclear whether quantum chaos is a necessary condition for RSB to occur. In order to further elucidate this issue, we study the, non quantum chaotic, $q = 2$ nonhermitian
SYK model which admits an explicit analytical solution of the Schwinger-Dyson equations. It can also be solved, see \cite{us2021} for details, by mapping it onto a model of free fermions. 
The free energy can be expressed, see also \cite{us2021}, as a sum over Matsubara frequencies.
 Each
  time a new Matsubara frequency enters the sum, by
  lowering the temperature,
  a  second order phase transition occurs. Kinks in $-dF/dT$, depicted in Fig.~\ref{fig:dfree}, indicate the positions of the critical temperatures. 
  The propagator $G_{LR}(\tau)$ can be also expressed as a finite sum over Matsubara frequencies so it does not depend
  exponentially on $\tau$. Because of the absence of a gap,
  there is no direct relation between RSB and wormholes which further suggests that the physics is qualitatively different from the quantum chaotic case.
  Further research is needed to delimit the
  importance of quantum chaos in RSB.

  In summary, we have provided evidence that RSB configurations dominate the low temperature phase of the partition function of pairs of random non-hermitian,  quantum chaotic systems whose sum is PT-symmetric. These field theory configurations mimic the contribution of wormholes in the gravitational path integral. In both cases, a first order transition occurs when replica symmetric saddle points take control of the partition function.

 \acknowledgments{
   This work was supported by NSFC Grant No. 11874259 (AMG), the National Key R$\&$D Program of China (Project ID: 2019YFA0308603) (AMG), a Shanghai talent program (AMG), NRF grant NRF-2020R1C1C1007591 (DR), the Institute for Basic Science in Korea (IBS-R024-D1) (DR) and U.S. DOE Grant
No. DE-FAG-88FR40388 (YJ and JJMV).
 We thank Juan Diego Urbina, Jeff Murugan, Tomoki Nosaka,
Klaus Richter and Zhenbin Yang for illuminating correspondence.
}

\bibliographystyle{apsrev4-1}
\bibliography{librarynh.bib}

\end{document}